\documentclass[12pt]{iopart}
\usepackage{iopams}
\usepackage{epsfig}

\def\bea{\begin{eqnarray}}  \def\eea{\end{eqnarray}}
\def\beq{\begin{equation}}   \def\eeq{\end{equation}}

\def\beeq{\begin{eqnarray}} \def\eeeq{\end{eqnarray}}

\begin{document}

\title[Multiplicities and $J/\psi$ suppression at LHC energies]{Multiplicities 
and $J/\psi$ suppression at LHC energies}

\author{A. Capella$^1$ and E. G. Ferreiro$^2$}

\address{$^1$
Laboratoire de Physique Th\'eorique,
Universit\'e de Paris XI, B\^atiment 210, 91405 Orsay Cedex, France}
\address{$^2$
Departamento de F{\'\i}sica de Part{\'\i}culas,
Universidad de Santiago de Compostela, 15782 Santiago de Compostela,
Spain}
\begin{abstract}
We present our predictions on multiplicities and $J/\psi$ suppression at LHC energies.
Our results take into account shadowing effects in the initial state and final state interactions 
with the hot medium. We obtain 1800 charged particles at LHC and 
the $J/\psi$ suppression increases by a factor 5 to 6 compared to RHIC. 
\end{abstract}

\section{Multiplicities with shadowing corrections}

%Multiplicities are usually considered as the addition of two contributions: one proportional to 
%number of
%participant
%nucleons $A$ %(valence-like contribution) 
%and a second one proportional to the number of 
%inelastic nucleon-nucleon collisions $A^{4/3}$, dominant at
%asymptotic energies.
%In order to get the right multiplicities at
%RHIC, d
At high energy, different mechanisms in the initial state {\it -shadowing-}, that lower the 
total multiplicity, have to be taken into account. 
%have been included to lower this second contribution.
The shadowing makes the
nuclear structure functions in nuclei different from the superposition of
those of their constituents nucleons. Its effect 
increases with decreasing $x$ and decreases with increasing
$Q^2$.
We have included a dynamical, non linear  of shadowing \cite{Capella:2005cn}, controlled by triple pomeron diagrams.
It is determined in terms of the diffractive cross sections. 
Our results for charged particles multiplicities at RHIC and LHC energies are presented in Fig. 1.
In absence of shadowing we obtain a maximal multiplicity, $dN_{AA}/dy = A^{4/3}$. With shadowing 
corrections the multiplicity behaves as $dN_{AA}/dy = A^{\alpha}$, with $\alpha=1.13$ at RHIC
and $\alpha=1.1$ at LHC.

\begin{center}
\begin{figure*}
\begin{minipage}[t]{80mm}
\epsfxsize=7.5cm
\epsfysize=8.5cm
\centerline{\epsfbox{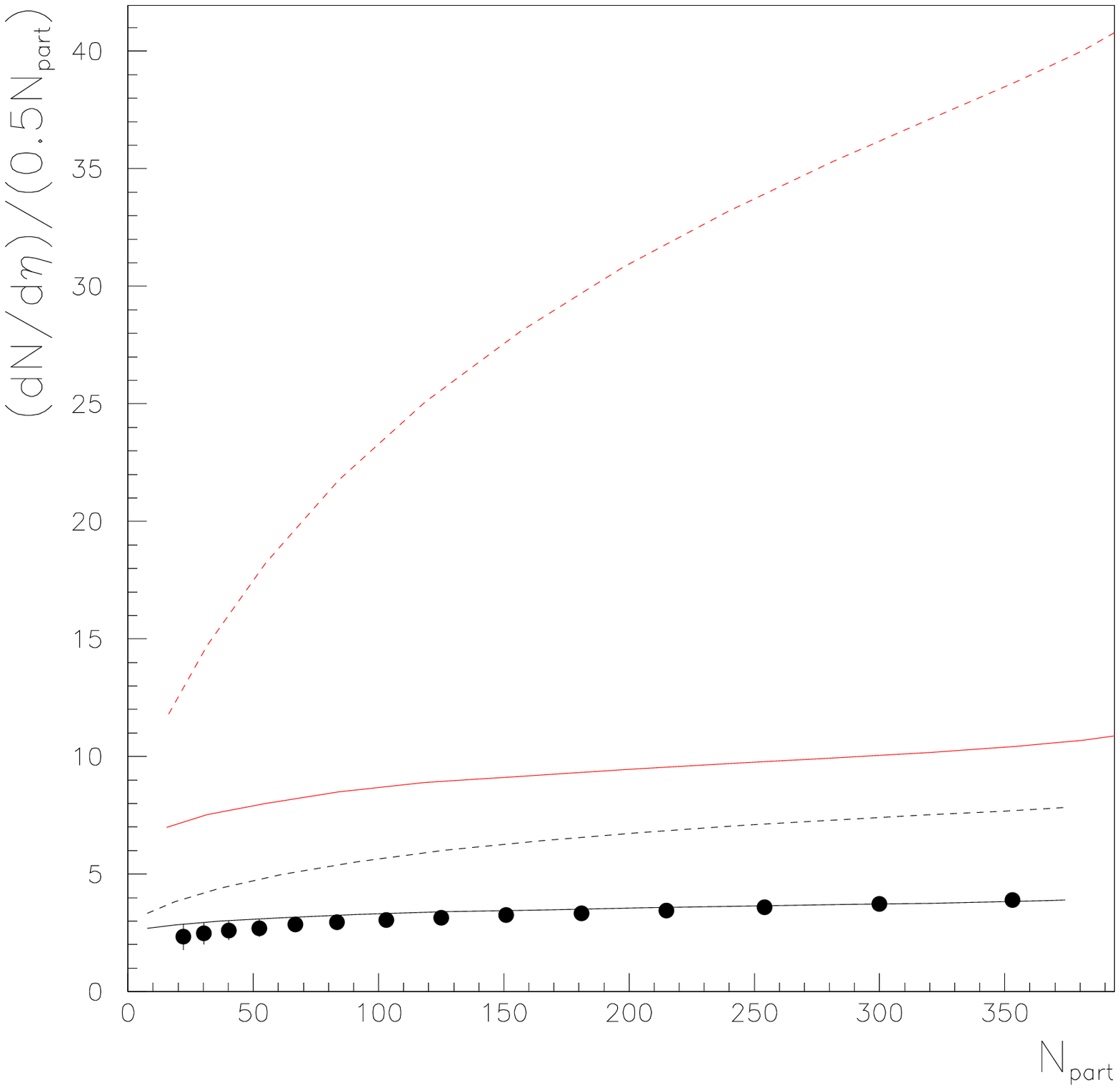}}
\vskip -1.85cm
\caption{
Multiplicities of charged particles with (solid lines) and without (dashed lines)
shadowing corrections at RHIC and LHC.}
\end{minipage}
\hspace{\fill}
\begin{minipage}[t]{80mm}
\vskip -8.5cm
\epsfxsize=7.5cm
\epsfysize=8.5cm
\centerline{\epsffile[4 4 500 650]{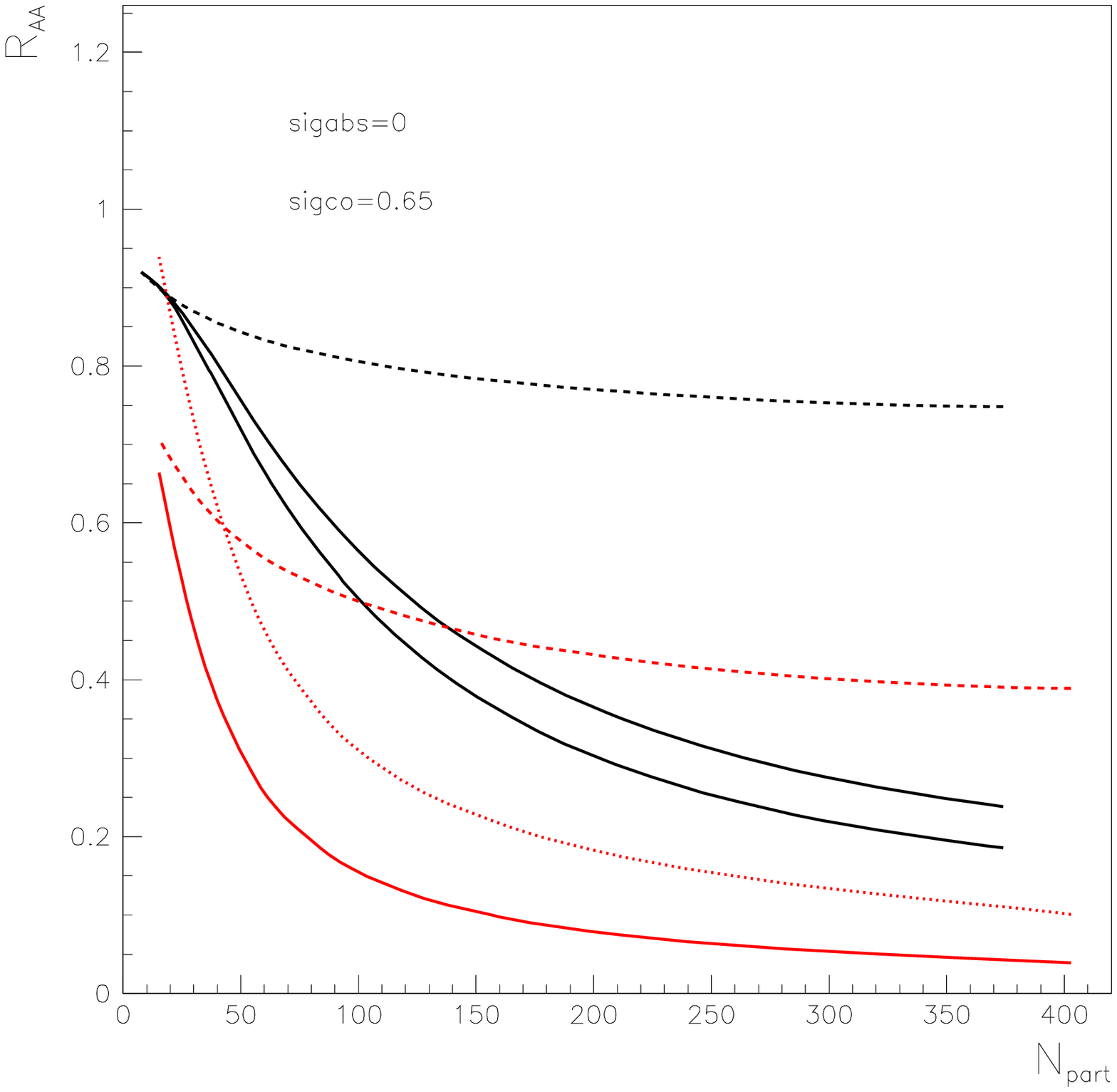}}
\vskip -1.85cm
\caption{$J/\psi$ production at RHIC and LHC. 
Dashed: $J/\psi$ shadowing, pointed: comovers suppression, continuous: total suppression.}
\end{minipage}
\vskip -0.3cm
\end{figure*}
\end{center}

\section{$J/\psi$ suppression}

%The $J/\psi$ production in proton-nucleus collisions
%is suppressed with
%respect to the characteristic $A^1$ scaling of lepton pair production.
%This
% suppression is generally
%interpreted as a
%result of the multiple scattering of a pre-resonance $c\overline{c}$
%with the nucleons of the nucleus: {\it nuclear absorption}.
An anomalous 
$J/\psi$ suppression 
-that clearly exceeds the one
expected from nuclear absorption-
has been found in $PbPb$ collisions at SPS.
Such a phenomenon was predicted by Matsui and Satz as a consequence
of
deconfinement in a dense medium.
It can also be described as a result
of final state interaction of the $c\overline{c}$ pair with the dense
medium produced in the collision: {\it comovers
interaction} \cite{Capella:2000zp}.
Here we present our results for the ratio of the $J/\psi$
yield over the average
number of binary nucleon-nucleon collisions at RHIC and LHC energies:
\beq
R_{AB}^{J/\psi}(b) = {dN_{AB}^{J/\psi}(b)/dy  \over n(b)} =
{dN_{pp}^{J/\psi}  \over dy}  {\int d^2s\ \sigma_{AB}(b)\ n(b,s) \
S^{abs}(b,s)\ S^{co}(b,s) \over \int d^2s\ \sigma_{AB}(b)\ n(b,s)}\ .
\eeq
$S^{abs}$ refers to the survival probability due to nuclear absorption and
$S^{co}$ is the survival probability due to the medium interactions.
The data on
$dAu$ collisions at RHIC favorize a small $\sigma_{abs}=0$ mb, so $S^{abs}=1$
\cite{Capella:2006mb}.
The interaction of a particle or a parton with the medium
is described by the
gain and loss differential equations which govern the final state
interactions:
\beq
\tau {d\rho^{J/\psi}(b,s,y) \over d \tau} = - \sigma_{co}\
\rho^{J/\psi}(b,s,y)\ \rho^{medium}(b,s,y)\ ,
\eeq
where $\rho^{J/\psi}$ and
$\rho^{co}=\rho^{medium}$ are the densities
of $J/\psi$ and comovers. 
We neglect a gain term resulting
from the recombination of $c$-$\overline{c}$ into $J/\psi$.
Our equations have to be integrated
between initial time $\tau_0$ and freeze-out time $\tau_f$. We use the inverse proportionality between proper time and densities,
$\tau_f/\tau_0 = \rho(b,s,y)/\rho_{pp}(y)$.
Our densities
 can be either hadrons or partons, so $\sigma_{co}$ represents an  effective cross-section
averaged over the interaction time.
We obtain the survival probability
$S_{co}(b,s)$
of the $J/\psi$ due to the medium interaction:
\begin{equation}
S^{co}(b,s) \equiv \frac{N^{J/\psi (final)}(b,s,y)}{N^{J/\psi
(initial)}(b,s,y)} \nonumber \\
= \exp \left [ - \sigma_{co}\ \rho^{co}(b,s,y) \ell n \left
(\frac{\rho^{co}(b,s,y)}{\rho_{pp}(0)} \right ) \right ] \ .
\end{equation}
The shadowing produces a decrease of the medium density. Because of this, the 
shadowing corrections on comovers increase
the $J/\psi$ survival probability $S^{co}$. On the other side, the shadowing corrections on ${J/\psi}$ 
decrease the $J/\psi$ yield.
Our results for RHIC and LHC are presented in Fig. 2.
We use the same
value of the comovers cross-section, $\sigma_{co} = 0.65$~mb that we have used at SPS energies.
We neglect the nuclear absorption.
%, $\sigma_{abs} = 0$~mb. 
The 
shadowing is introduced in both the comovers and the $J/\psi$ yields.

\end{document}